# Properties of MgB$_2$ thin films with carbon doping

A. V. Pogrebnyakov[a)], X. X. Xi

*Department of Physics, Department of Materials Science and Engineering,
and Materials Research Institute,
The Pennsylvania State University, University Park, PA 16802*

J. M. Redwing, V. Vaithyanathan, D. G. Schlom, A. Soukiassian

*Department of Materials Science and Engineering and Materials Research Institute,
The Pennsylvania State University, University Park, PA 16802*

S. B. Mi, C. L. Jia

*Institut für Festkőrperforschung,
Forschungszentrum Jűlich GmbH, D-52425 Jűlich, Germany*

J.E. Giencke, C. B. Eom

*Department of Materials Science and Engineering and
Applied Superconductivity Center,
University of Wisconsin, Madison, Wisconsin 53706, USA*

J. Chen, Y. F. Hu, Y. Cui, Q. Li

*Department of Physics and Materials Research Institute,
The Pennsylvania State University, University Park, PA 16802*

## Abstract

We have studied structural and superconducting properties of MgB$_2$ thin films doped with carbon during the hybrid physical-chemical vapor deposition process. A carbon-containing metalorganic precursor bis(methylcyclopentadienyl)magnesium was added to the carrier gas to achieve carbon doping. As the amount of carbon in the film increases, the resistivity increases, $T_c$ decreases, and the upper critical field increases dramatically as compared to clean films. The self-field $J_c$ in the carbon doped film is lower than that in the clean film, but $J_c$ remains relatively high to much higher magnetic fields, indicating stronger pinning. Structurally, the doped films are textured with columnar nano-grains and highly resistive amorphous areas at the grain boundaries. The carbon doping approach can be used to produce MgB$_2$ materials for high magnetic-field applications.

[a)]Electronic address: avp11@psu.edu

The 39-K superconductor $MgB_2$ [1] has attracted tremendous interest for its potential in high magnetic-field applications. In particular, high critical current density $J_c$ near the depairing limit has been observed in $MgB_2$ [2], and unlike high temperature superconductors grain boundaries in $MgB_2$ do not behave like weak links [3]. Although clean $MgB_2$ shows low upper critical field $H_{c2}$ [4], in high resistivity $MgB_2$ films $H_{c2}$ is substantially higher [5]; the critical current density $J_c$ in clean $MgB_2$ is suppressed by a relatively low magnetic field [6], but defects and impurities have been shown to enhance pinning [7, 8]. Recently, we have developed a hybrid physical-chemical vapor deposition (HPCVD) technique, which produces *in situ* epitaxial $MgB_2$ films with $T_c$ above 40 K [6, 10]. Because of the highly reducing $H_2$ ambient and the high purity $B_2H_6$ source used in the HPCVD process, the technique produces very clean $MgB_2$ thin films with a residual resistivity at $T_c$ as low as 0.28 µΩ.cm [10]. These clean films show low $H_{c2}$ values. In this letter, we describe the properties of HPCVD $MgB_2$ films doped with carbon. The $H_{c2}$ values in these ``dirtier'' films are significantly higher and the vortex pinning is significantly stronger than those in the clean films. Structural analyses of the carbon-doped $MgB_2$ films by x-ray diffraction and transmission electron microscopy (TEM) are presented.

*In situ* epitaxial growth of $MgB_2$ films by HPCVD have been described in detail previously [9]. Briefly, pure magnesium slugs are placed around the substrate on the top surface of a susceptor in a vertical quartz reactor. Hydrogen is used as the carrier gas. When the susceptor is heated inductively to around 720-760°C, a high Mg vapor pressure is generated near the substrate. The boron precursor gas, 1000 ppm diborane ($B_2H_6$) in $H_2$, is then introduced into the reactor and the $MgB_2$ film growth begins. The total pressure in the reactor during the deposition is 100 Torr. In the present work, the flow rate of the $H_2$ carrier gas was kept at 300 sccm and that of the $B_2H_6$ gas mixture at 150 sccm. The films were deposited on (001) 4H-SiC substrates at 720°C. The thickness of the films was around 2000 Å.

For carbon doping, we added bis(methylcyclopentadienyl)magnesium (($C_6H_7)_2Mg$ or $(MeCp)_2Mg$), a metalorganic magnesium precursor, to the carrier gas. A secondary hydrogen flow was passed through the $(MeCp)_2Mg$ bubbler which was held at 760 Torr and 21.6°C to transport $(MeCp)_2Mg$ to the reactor. The $(MeCp)_2Mg$ molecules decompose upon reaching the heated substrate, and the carbon-containing ligands in the precursor react with other species to dope the $MgB_2$ films. The amount of carbon doping depends on the secondary hydrogen flow rate through the $(MeCp)_2Mg$ bubbler, which was varied between 25 and 200 sccm to vary the flow rate of $(MeCp)_2Mg$ into the reactor from 0.0065 to 0.052 sccm. The higher the flow rate of $(MeCp)_2Mg$, the higher the carbon doping level in the $MgB_2$ films. The chemical compositions of a series of films were determined by wavelength dispersive X-ray spectroscopy (WDS) to establish a correlation between the carbon concentrations in the films and the hydrogen flow rates through the $(MeCp)_2Mg$ bubbler. For example, hydrogen flow rates of 50, 100, 150 sccm through the bubbler correspond to 15, 29, and 39 at.% of carbon in the films, respectively. The nominal atomic concentrations determined by this approach are used as the carbon concentrations presented in this paper.

The resistivity (in log scale) versus temperature curves for $MgB_2$ films with different carbon doping levels are shown in Fig. 1 (a). The carbon doping causes a dramatic increase in the resistivity, whereas the $T_c$ of the film is suppressed much more slowly. For example, with a carbon concentration of 24 at.%, the residual resistivity increases from the undoped value of less than 1 $\mu\Omega.cm$ to ~ 200 $\mu\Omega.cm$, but $T_c$ only decreases from over 41 K to 35 K. The dependences of residual resistivity and $T_c$ on the carbon concentration in the doped $MgB_2$ films are plotted in Fig. 1(b). $T_c$ is suppressed to below 4.2 K at a nominal carbon concentration of 42 at.% when the residual resistivity is 440 $m\Omega.cm$. This is very different from those in carbon doped single crystals, where $T_c$ is suppressed to 2.5 K at a residual resistivity of 50 $\mu\Omega.cm$ when 12.5 at.% of carbon is doped into $MgB_2$ [11]. This discrepancy indicates that only a small portion of the carbon in the films is doped into the $MgB_2$ structure and the rest most likely forms high resistance grain boundaries giving rise to poor connectivity of the $Mg(B_{1-x}C_x)_2$ grains [12].

The granular structure of the carbon-doped $MgB_2$ films is confirmed by TEM. Figure 2(a) is a cross-sectional TEM image of a film with 22 at.% nominal carbon concentration taken along the [$\bar{1}10$] direction of the substrate. It shows that the film consists of columnar nano-grains (the contrast changes laterally, but not vertically) of $Mg(B_{1-x}C_x)_2$ with a preferential $c$-axis orientation. The selected area electron diffraction pattern taken from the $MgB_2$/SiC interface area in Fig. 2(b) shows two types of features, diffraction spots and arcs. The spots belong to the single crystal SiC substrate (SC) and the arcs to the $MgB_2$ film (MB). The arcs consist of many fine spots originating from individual columnar grains, which show a deviation of their $c$-axis from the film normal. In the planar-view image in Fig. 2(c), the change of contrast indicates an equiaxial in-plane morphology of the columnar grains, and an amorphous phase is also observed between the grains. We were not able to determine the composition of the amorphous areas, but it is most likely that the large portion of carbon that is not doped into $MgB_2$ is contained in these areas. The insert in Fig. 2(c) is a typical diffraction pattern taken along the film normal. The strong hexagonal-distributed spots show that the hexagonal-on-hexagonal in-plane relationship between the columnar grains and SiC dominates, while the diffraction rings reveal grains that are randomly in-plane oriented.

Figure 3(a) shows $\theta$-$2\theta$ scans of an undoped $MgB_2$ film and films doped with different amounts of carbon. Compared to the undoped films, the $MgB_2$ 00$l$ peaks are suppressed as carbon concentration increases, and decrease dramatically when the carbon concentration is above ~ 30 at.%. Meanwhile, as shown in Fig. 3(b), both the $c$ and $a$ axes expand until about 30 at.%, above which the $c$ lattice constant decreases and the $a$ lattice constant increases dramatically. This behavior is different from that in carbon-doped single crystals, where the $a$ axis lattice constant decreases but that of *c* axis remains almost constant for all the carbon concentration [11]. The peak marked by ``?'' is likely 101 MgB$_2$, the most intense diffraction peak of MgB$_2$. It becomes stronger as the carbon concentration increases, indicating an increased presence of randomly oriented MgB$_2$. The peaks marked by ``Boron Carbide'', according to extensive pole figure analysis, are most likely due to B$_4$C, B$_8$C, or B$_{13}$C$_2$. Their intensities also increase with carbon concentration. From the TEM and x-ray diffraction results, we conclude that below about 30 at.%, a small portion of carbon is doped into the Mg(B$_{1-x}$C$_x$)$_2$ columnar, *c*-axis-oriented nano-grains, which have larger *c* and *a* lattice constants than in the undoped MgB$_2$. The rest of carbon goes into the grain boundaries consisting of highly resistive amorphous phases or boron carbides. Above about 30 at.%, the Mg(B$_{1-x}$C$_x$)$_2$ nano-grains are completely separated from each other by highly resistive phases, become more randomly oriented, and their lattice constants relax. The resistivity versus carbon doping result in Fig. 1(b) is consistent with this scenario.

The upper critical field $H_{c2}$ was measured using a Quantum Design PPMS system with a 9-T superconducting magnet. Figure 4 shows the results for an undoped, 7.4 at.%, and 22 at.% carbon doped films. The value of $H_{c2}$ is defined by 50% of the normal-state resistance $R(H_{c2}) = 0.5R(T_c)$. It can be clearly seen that carbon doping changes the downward curvature in $H_{c2}^\perp(T)$ for the undoped film to an upward curvature in the carbon doped films. Both the slope, $dH_{c2}/dT$, near $T_c$ and the low temperature $H_{c2}$ increase with carbon concentration, even when the $T_c$ of the film is reduced. In high magnetic field measurements, Braccini *et al*. have shown that carbon doped MgB$_2$ films as described here have extraordinary $H_{c2}(0)$ values as high as 70 T [13]. The transport $J_c(H)$ at different temperatures for an undoped and a carbon doped MgB$_2$ films are shown in Fig. 5. Bridges of 20 – 50 µm in width were patterned in the films by photolithography and Ar ion beam milling, and a criterion of 1 µV was used to determine $J_c$ from the *I-V* curves. While the undoped film has high self-field critical current densities, they are suppressed quickly by magnetic field due to the weak pinning (see Fig. 5(a)). For the film doped with 11 at.% nominal carbon concentration, Fig. 5(b), the self-field $J_c$ is lower than in the clean film, which is expected due to reduced connectivity between the Mg(B$_{1-x}$C$_x$)$_2$ grains [12], but $J_c$ values are relatively high in much higher magnetic fields. This indicates a significantly enhanced vortex pinning in carbon doped MgB$_2$ films.

In conclusion, MgB$_2$ thin films were doped with carbon by adding bis(cyclopentadienyl)magnesium to the carrier gas during the HPCVD process. The residual resistivity increases rapidly while $T_c$ decreases much more slowly with carbon doping. Structural analyses show that only part of the carbon is doped into the MgB$_2$ lattice and the rest forms highly resistive foreign phases in the grain boundaries. $H_{c2}(T)$ and its slope near $T_c$ increase dramatically as compared to the clean films, which is consistent with the multiband superconductor model of Gurevich and indicates a dirtier $\pi$ band upon carbon doping [14]. The critical current density in magnetic field increases markedly from that in the clean film due to stronger vortex pinning. The results demonstrate that the carbon doping approach may be used to enhance $H_{c2}$ and $J_c(H)$ of $MgB_2$ to make practical materials for high magnetic-field applications.

The work at Penn State is supported in part by ONR under grant Nos. N00014-00-1-0294 (Xi) and N0014-01-1-0006 (Redwing), by NSF under grant Nos. DMR-0306746 (Xi and Redwing), DMR-9876266 and DMR-9972973 (Li), and by DOE under grant No. DE-FG02-03ER46063 (Schlom). The work at University of Wisconsin is supported by the National Science foundation through the MRSEC for Nanostructure Materials.

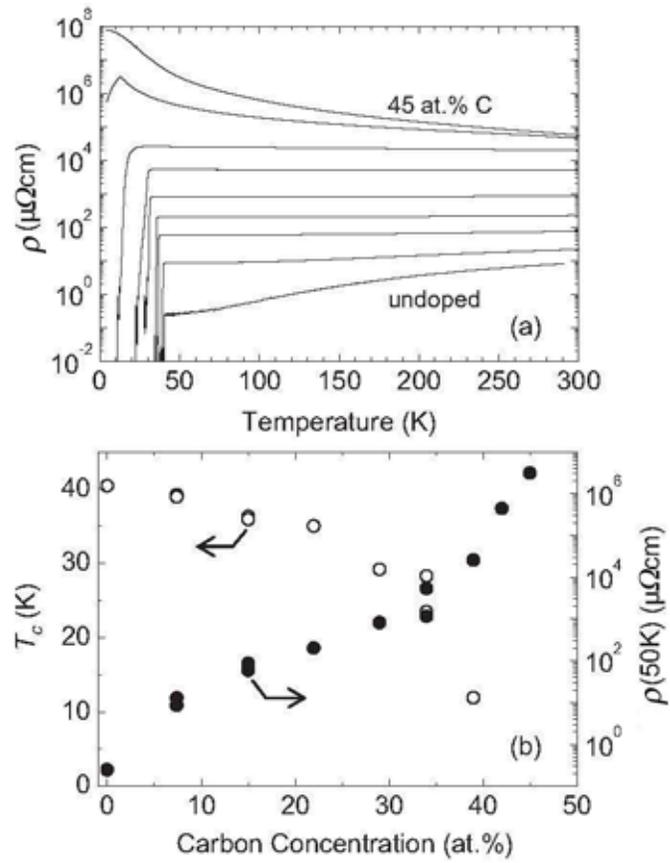

Fig. 1. (a) Resistivity versus temperature curves for MgB$_2$ films of different carbon doping. (b) Residual resistivity (solid circles) and $T_c$ (open circles) as a function of carbon concentration for films plotted in (a). In (a), from bottom to top, the nominal carbon concentrations of the curves are 0, 7.4, 15, 22, 29, 34, 39, 42, and 45 at.%.

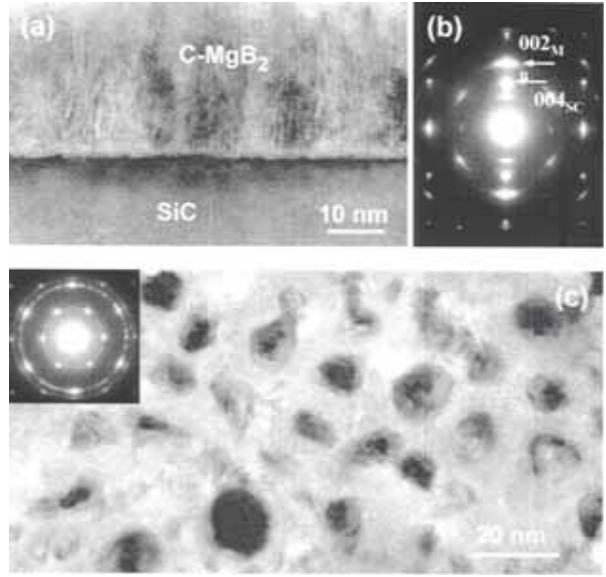

Fig. 2. TEM study results of a film with 22 at.% nominal carbon concentration. (a) Cross-sectional image taken along the [$\bar{1}$10] direction of the substrate. (b) Selected area electron diffraction taken from the MgB$_2$/SiC interface area. (c) Planar-view image showing nano-grains of carbon doped MgB$_2$ and an amorphous phase between the grains. The insert is the select area electron diffraction pattern taken along the film normal.

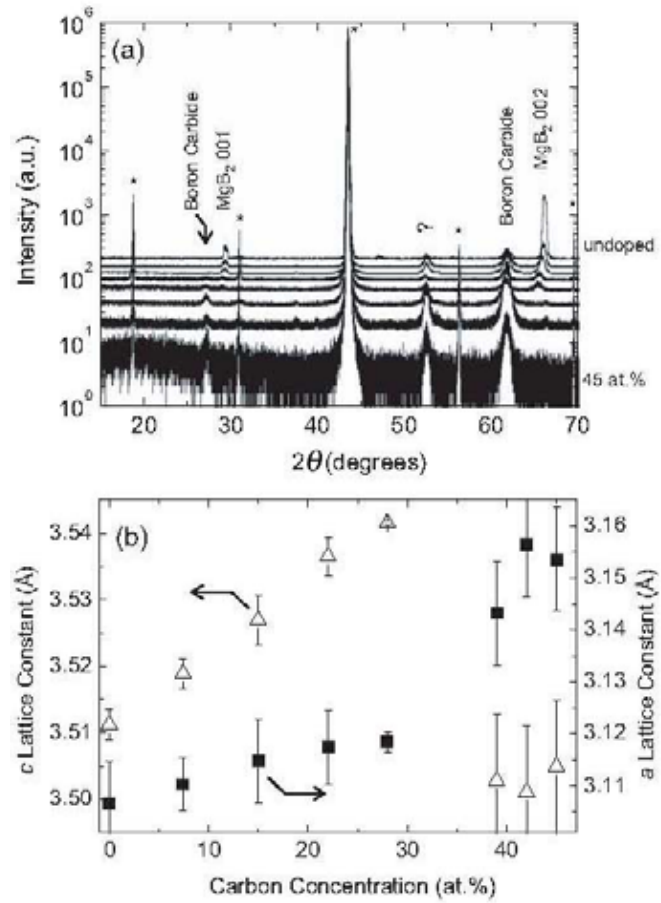

Fig. 3. (a) X-ray diffraction $\theta$-$2\theta$ scans for $MgB_2$ films with carbon doping. From top to bottom, the nominal carbon concentrations are 0, 7.4, 15, 22, 28, 39, 42, and 45 at.%. The spectra are shifted vertically for clarity. The peaks labeled with ``*'' are due to the SiC substrate peaks. (c) The $c$-axis lattice constant (open triangles) and $a$-axis lattice constant (solid squares) of the carbon doped $MgB_2$ films as a function of nominal carbon concentration.

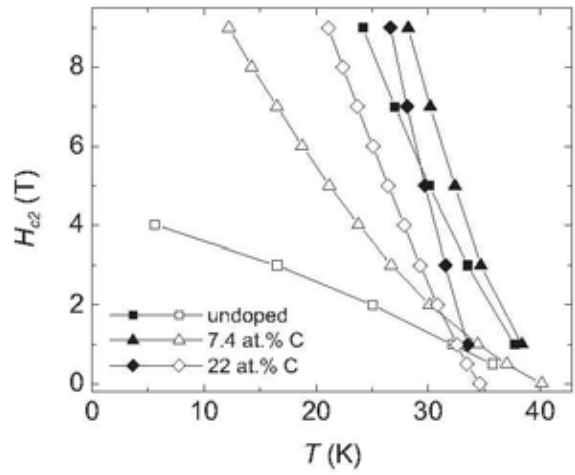

Fig. 4. Upper critical field as a function of temperature for an undoped film and two doped films with 7.4 at.% and 22 at.% nominal carbon concentrations, respectively. The solid symbols are for parallel field ($H_{c2}^{\parallel}$) and the open symbols are for perpendicular field ($H_{c2}^{\perp}$).

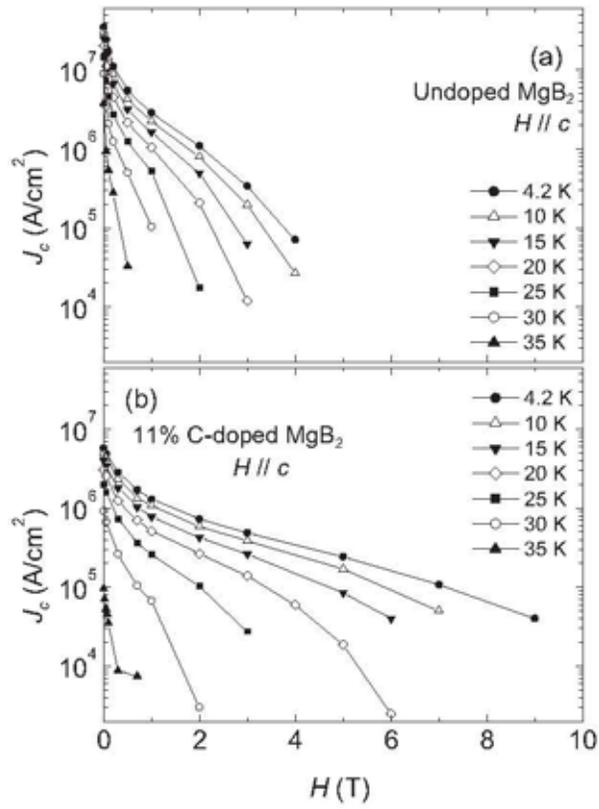

Fig. 5. Critical current density as a function of magnetic field ($\mathbf{H} \parallel c$) and temperature for (a) an undoped film and (b) a film doped with with 11 at.% nominal carbon concentration.